%% file: main.tex
\documentclass[conference]{IEEEtran}
\IEEEoverridecommandlockouts
\usepackage{cite}
\usepackage{amsmath,amssymb,amsfonts}
\usepackage{algorithmic}
\usepackage{graphicx}
\usepackage{subcaption}
\usepackage{textcomp}
\usepackage{booktabs}
\usepackage[table,xcdraw]{xcolor}
\usepackage{caption}
\usepackage{svg}
\usepackage{pifont}
\def\BibTeX{{\rm B\kern-.05em{\sc i\kern-.025em b}\kern-.08em
    T\kern-.1667em\lower.7ex\hbox{E}\kern-.125emX}}

\usepackage{hyperref}
\hypersetup{hypertex=true,
colorlinks=true,
linkcolor=[RGB]{105,33,106},
anchorcolor=[RGB]{105,33,106},
citecolor=[RGB]{105,33,106},
urlcolor=[RGB]{0,76,129}}

\usepackage[ruled,vlined]{algorithm2e}
\usepackage{makecell}
\usepackage{multirow}
\usepackage{verbatim}

\SetKwInput{KwData}{Data}
\SetKwInput{KwResult}{Result}

\newcommand{\method}{H\textsuperscript{2}EAL}

\begin{document}

\title{
    \huge 
    \textbf{\method: Hybrid-Bonding Architecture with Hybrid Sparse Attention for Efficient Long-Context LLM Inference} \\
    }

\author{%
  \IEEEauthorblockN{%
    Zizhuo Fu\textsuperscript{1,2,3\dag},\;
    Xiaotian Guo\textsuperscript{2\dag},\;
    Wenxuan Zeng\textsuperscript{1},\;
    Shuzhang Zhong\textsuperscript{1,2},\\
    Yadong Zhang\textsuperscript{5},\;
    Peiyu Chen\textsuperscript{5},\;
    Runsheng Wang\textsuperscript{2},\;
    Le Ye\textsuperscript{2,4*},\;
    Meng Li\textsuperscript{1,2*}%
  }\\[-2ex]
  \IEEEauthorblockA{%
    \textsuperscript{1}Institute for Artificial Intelligence, 
    \textsuperscript{2}School of Integrated Circuits, Peking University, Beijing, China\\
    \textsuperscript{3}School of Electronics Engineering and Computer Science, Peking University, Beijing, China\\
    \textsuperscript{4}Advanced Institute of Information Technology of Peking University, Hangzhou, China\\
    \textsuperscript{5}Nano Core Chip Electronic Technology, Hangzhou, China
  }%
}



\maketitle

\captionsetup[table]{
  labelsep = colon,        
  justification = raggedright,  
  singlelinecheck = off,   
  labelfont = bf,         
  textfont = normalfont,   
  name = Table,           
  skip = 5pt              
}

\input{docs/0-abstract}

\input{docs/1-introduction}
\input{docs/2-background}

\input{docs/3-motivation}
\input{docs/4-method}

\input{docs/5-experiments}
\input{docs/6-conclusion}
\input{docs/7-acknowledgment}

\clearpage
\bibliographystyle{IEEEtran}
\bibliography{main}


\end{document}

%% file: docs/0-abstract.tex
\begin{abstract}
    Large language models (LLMs) have demonstrated remarkable proficiency in a wide range of natural language processing applications. 
    However, the high energy and latency overhead induced by the KV cache limits the edge deployment, especially for long contexts.
    Emerging hybrid bonding (HB) technology has been proposed as a promising alternative to conventional near-memory processing (NMP) architectures, offering improved bandwidth efficiency and lower power consumption while exhibiting characteristics of distributed memory.

    In this paper, we propose \method, an HB-based accelerator with sparse attention algorithm-hardware co-design for efficient LLM inference at the edge.
    \textit{At the algorithm level,} we propose a hybrid sparse attention scheme with static and dynamic sparsity for different heads to fully leverage the sparsity with high accuracy.
    \textit{At the hardware level,} we co-design the hardware to support hybrid sparse attention and propose memory-compute co-placement to address the distributed memory bottleneck.
    Since different attention heads exhibit different sparse patterns and the attention structure often mismatches the HB architecture, we further develop a load-balancing scheduler with parallel tiled attention to address workload imbalance and optimize the mapping strategy.
    Extensive experiments demonstrate \method~achieves $\text{5.20}\sim\text{48.21}\times$ speedup and $\text{6.22}\sim\text{73.48}\times$ energy efficiency improvement over baseline HB implementation, with a negligible average accuracy drop of 0.87\% on multiple benchmarks.





    
\end{abstract}




%% file: docs/1-introduction.tex
\section{Introduction}

\begin{figure}[!tb]
    \centering
    \includegraphics[width=1\linewidth]{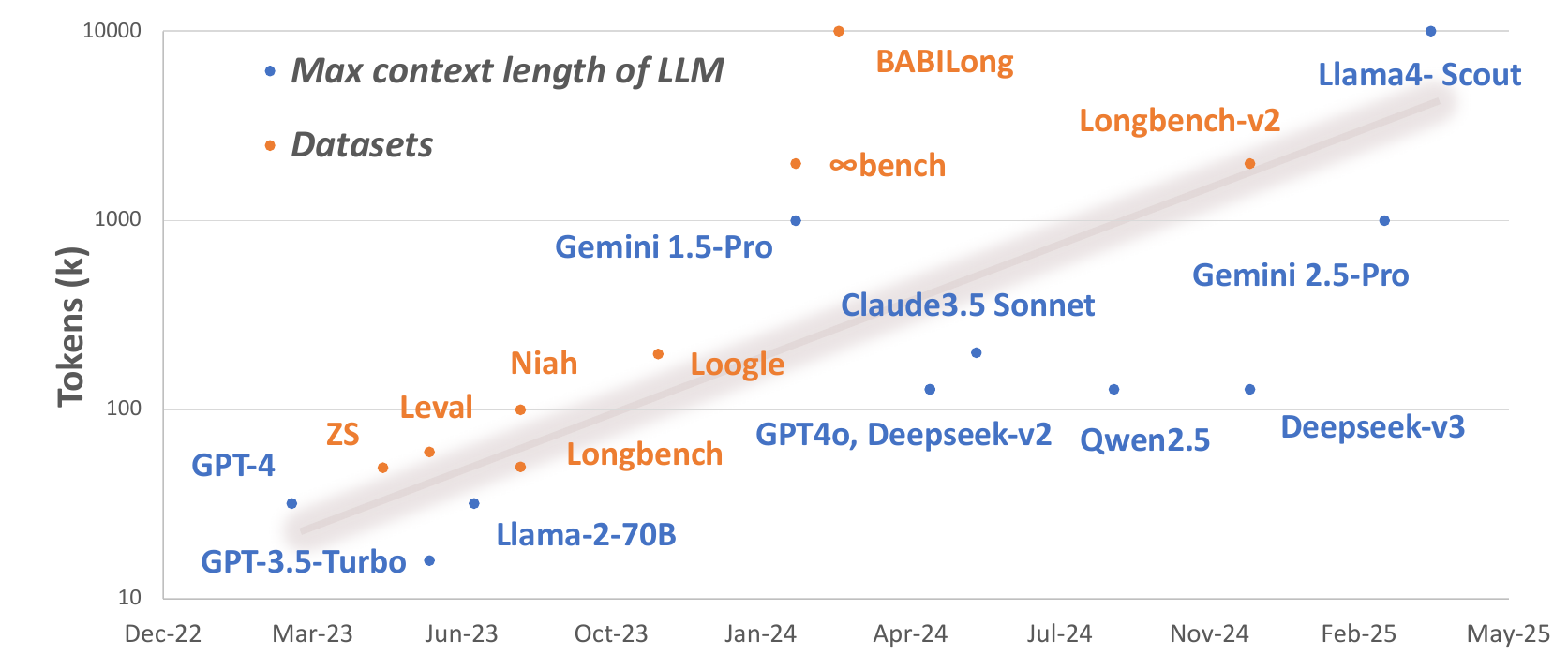}
    \caption{Trends in long-context processing, including the evolution of maximum context length of LLMs and datasets.
    }
    \label{fig:longcontext}
    \vspace{-0.4cm}
\end{figure}

Large language models (LLMs) have revolutionized various domains in field of generative artificial intelligence (AI) \cite{Yi2024Survey}.
While deployment on the cloud remains prevalent for mass user access, emerging trends show LLMs are deployed on intelligent edge platforms to address growing demands for localized functionality, personalized interactions, and data privacy \cite{Durante2024AgentAI,Huang2022LanguageModels}.
Edge-oriented LLM services prioritize small-batch processing while addressing growing demands for long-context applications such as long document analysis \cite{Koh2022AnES}, multi-turn dialogue \cite{Zhang2025ASO}, long chain-of-thought reasoning \cite{Chen2025TowardsRE,Qiu2025PHYBenchHE}. 
Figure \ref{fig:longcontext} demonstrates the trend in long-context processing and highlights hardware challenges in KV cache management.


Efficient LLM inference relies on the hardware to process two phases, i.e., prefill and decoding.
During the prefill phase, LLM processes the input prompt, leveraging parallel processing to process multiple tokens. 
In contrast, during the decoding phase, LLM follows auto-regressive generation and only outputs one token at each step, which is more dependent on the memory due to the low data reuse.
To meet both demands, near-memory processing (NMP)-based heterogeneous architectures have been proposed \cite{Heo2024NeuPIMsNH,Yun2024DuplexAD,Park2024AttAccUT}. 
NMP combines standard computing units with memory-integrated accelerators to improve both compute-intensive and memory-heavy operations.
However, traditional NMP approaches with in-die DRAM integration suffer from limited logic density and storage capacity \cite{Devaux2019TheTP}.
These challenges lead to lower throughput of attention computation and less effective key-value (KV) cache management as context lengths increase, ultimately affecting real-time performance in scenarios with long contexts and low batch sizes.


\begin{figure*}
    \centering
    \includegraphics[width=1\linewidth]{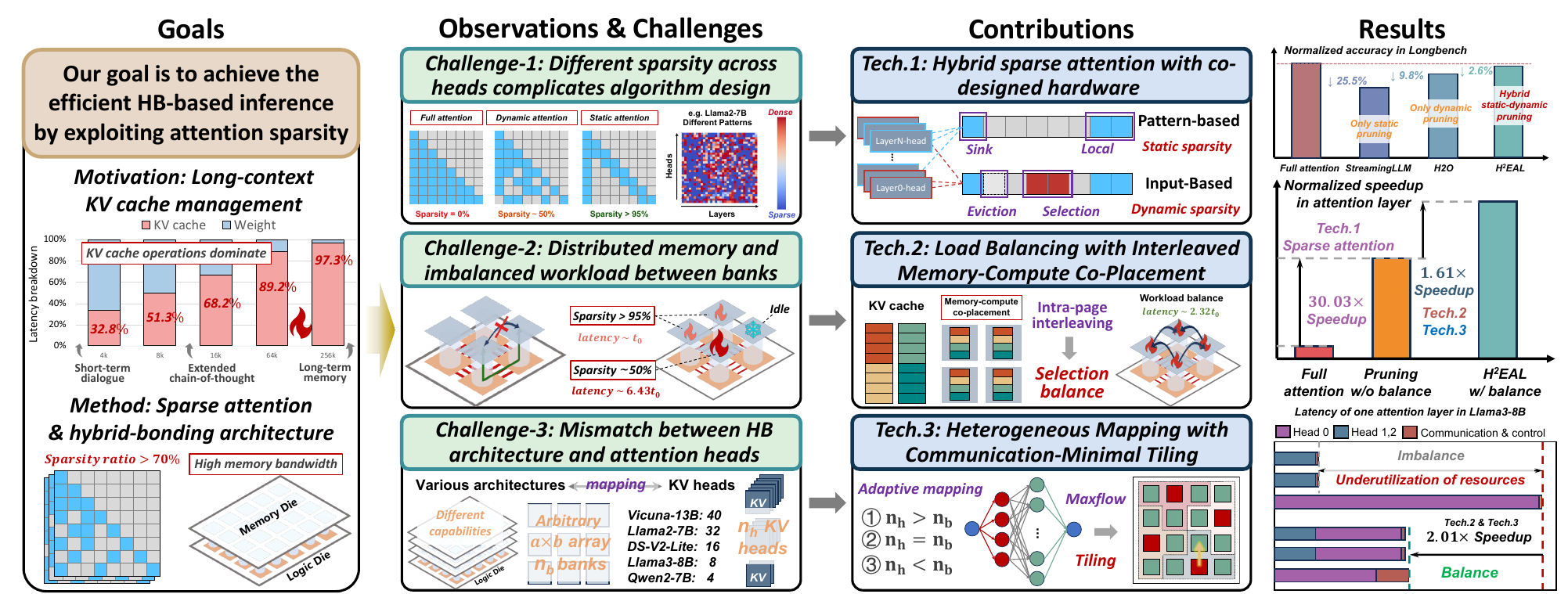}
    \caption{Overview of the challenges and contributions in \method.}
    \label{fig:overview}
\end{figure*}



The recent emerging hybrid bonding (HB) technology provides a promising alternative to conventional NMP architectures \cite{Yue2024ExploitingSO,Niu2022184QPSW6L,Fujun2020ASE,Wuu20223DVT,h2llm}.
It not only delivers substantial bandwidth with lower power consumption than HBM\cite{Fujun2020ASE,Niu2022184QPSW6L}, but also boosts NMP’s computational capacity by allowing processing engines to be implemented on the vertically stacked logic die.
However, it exhibits distinctive features that differentiate it from other architectures.
HB adopts a distributed-memory design, where the banks in memory dies are not directly interconnected, but connected to logic banks of high bandwidth with separate I/O and control units. 
This means that each bank operates largely independently, with a relatively limited storage capacity per bank.




Meanwhile, recent advancements in sparse attention methods \cite{Zhang2023H2OHO, Tang2024QuestQS} have shown strong potential for edge inference in long-context scenarios. 
These approaches significantly reduce KV cache overhead with minimal impact on the model accuracy, becoming particularly beneficial for resource-constrained edge environments. 
However, previous methods are primarily designed for GPU-like systems, which rely on unified memory and enable seamless data sharing across all processing cores. 
These methods are not suitable for HB architectures due to the distributed memory. 
Additionally, different attention sparsity across heads has been observed in \cite{Xiao2024DuoAttentionEL,Fu2024NotAH}.
Therefore, allocating a unified sparse pattern for all heads is insufficient to fully leverage the sparsity.

To tackle these challenges, we propose \method, an HB-based accelerator that leverages hybrid static-dynamic sparse attention with algorithm-hardware co-design for efficient edge LLM inference. 
The main contributions can be summarized as
\begin{itemize}
    \item \textbf{Hybrid sparse attention with co-designed hardware to fully leverage the attention sparsity}: We propose an algorithm-hardware co-designed framework named \method. At the algorithm level, we propose a hybrid sparse attention mechanism, considering diverse sparse patterns across different heads. At the hardware level, we design the hardware based on the characteristics of HB architecture.
    \item \textbf{Load balancing with interleaved memory-compute co-placement}: To eliminate the workload imbalance between heads induced by hybrid sparse attention, we design a distributed load balancing approach that features interleaved KV cache storage and memory-compute co-placement KV cache operation.
    \item \textbf{Adaptive mismatch-aware heterogeneous mapping with communication-minimal tiling}: We develop a mismatch-aware mapping algorithm to automatically adapt arbitrary attention structures and the number of heads to HB architectures with different array configurations, achieving better resource utilization.
\end{itemize}

Extensive experiments demonstrate that \method~achieves $\text{5.20}\sim\text{48.21}\times$ speedup and $\text{6.22}\sim\text{73.48}\times$ energy efficiency improvement over baseline HB implementation with only a negligible accuracy degradation of 0.87\(\%\) on average.

%% file: docs/2-background.tex
\section{Background}

\subsection{HB Architecture for Edge-side LLM Inference}

Deploying LLMs on edge devices for smart homes, autonomous driving, and IoT demands small-scale, real-time tasks to meet latency and power constraints \cite{Durante2024AgentAI, Huang2022LanguageModels}. 
For long-context LLM inference where models handle extended lengths of sequence, edge devices face a unique challenge since the KV cache size becomes significant \cite{Luohe2024KVCache}.
Storing and accessing large KV cache require high memory bandwidth demands, which are difficult for traditional memory architectures, especially for edge devices with limited resources. 

Emerging HB technology presents a promising solution by vertically stacking memory dies (e.g., DRAM) on top of logic dies and interconnecting them through high-bandwidth copper (Cu-Cu) direct bonding \cite{Yue2024ExploitingSO,Niu2022184QPSW6L,Fujun2020ASE,Wuu20223DVT} as illustrated in Figure \ref{fig:HBLLM}(a). 
This 3D integration enables substantial bandwidth and low power consumption, making HB a promising alternative to HBM and in-die NMP. However, despite its advantages, HB still faces challenges in managing memory and compute resources that are distributed spatially, especially for long-context inference where the KV cache becomes the dominant. 
Balancing the computational demands of LLMs with the memory bandwidth available in HB architecture is crucial for edge-side performance.

\subsection{Sparse Attention Algorithm}

Long-context LLM inference increases the computation and memory requirements induced by the KV cache, which becomes the prominent bottleneck.
To alleviate this, sparse attention techniques have been proposed to reduce the number of attended tokens during each decoding step.
These methods can be divided into \textbf{static sparse attention (pattern-based)} and \textbf{dynamic sparse attention (input-based)}. 
Static sparsity such as the first few or local tokens uses pre-defined attention patterns to limit the number of tokens retained in the KV cache, which directly reduces memory usage but may risk losing important context \cite{Xiao2023EfficientSL,Ge2023ModelTY}. 
In contrast, dynamic sparsity prunes less relevant tokens according to the input-based attention scores, offering better flexibility and utilization of the memory \cite{Zhang2023H2OHO,Tang2024QuestQS,Chen2024ArkValeEG}.

Hardware-based sparse attention has been explored to reduce memory usage and computational overhead in LLMs \cite{Tu2023TranCIMFB, sadimm, Zhang2023H2OHO}.  
TranCIM \cite{Tu2023TranCIMFB} adopts fixed-pattern sparsity based on the StreamingLLM algorithm \cite{Xiao2023EfficientSL}, restricting attention to local tokens and reducing data movement but yielding inflexible, suboptimal long-context performance.  
SADIMM \cite{sadimm} employs top-k attention-based dynamic sparsity, improving adaptability but incurring hardware overhead for token ranking and risking accuracy loss\cite{Tang2024QuestQS, Chen2024ArkValeEG}.  
H2O \cite{Zhang2023H2OHO} evicts low-importance tokens yet underuses attention sparsity\cite{Xiao2024DuoAttentionEL, Yang2025LServeEL},  and remains bottlenecked by GPU memory bandwidth during decoding.
As shown in Table \ref{tab:comparison}, these studies underscore the challenges in balancing sparse attention efficiency with hardware complexity, latency, and power consumption.

\begin{table}[]
\centering
\caption{Qualitative comparison of proposed \method~with state-of-the-art LLM acceleration works with sparse attention.}
\label{tab:comparison}

\resizebox{\columnwidth}{!}{
\begin{tabular}{@{}ccccc@{}}
\toprule
\textbf{Method}                           & TranCIM\cite{Tu2023TranCIMFB}                                                       & SADIMM\cite{sadimm}                                                          & H2O\cite{Zhang2023H2OHO}                                                      & {\color[HTML]{0D0EFF} \textbf{\method~(This work)}}                                         \\ 
\specialrule{0.1em}{3pt}{3pt}
\cellcolor[HTML]{F1F1F1}\textbf{Hardware platform}         & SRAM CIM                                                      & DIMM-based NMP                                                  & GPU                                                      & \textbf{Hybrid-bonding}                                                             \\
\cmidrule(lr){2-5}
\cellcolor[HTML]{F1F1F1}\textbf{Static sparsity}    & \begin{tabular}[c]{@{}c@{}}Yes\\ (Fixed pattern)\end{tabular} & No                                                              & No                                                       & \textbf{\begin{tabular}[c]{@{}c@{}}Yes\\ (Sink + local)\end{tabular}}               \\
\cmidrule(lr){2-5}
\cellcolor[HTML]{F1F1F1}\textbf{Dynamic sparsity}   & No                                                            & \begin{tabular}[c]{@{}c@{}}Yes\\ (Top-k selection)\end{tabular} & \begin{tabular}[c]{@{}c@{}}Yes\\ (Eviction)\end{tabular} & \textbf{\begin{tabular}[c]{@{}c@{}}Yes\\ (Top-k selection + eviction)\end{tabular}} \\
\cmidrule(lr){2-5}
\cellcolor[HTML]{F1F1F1}\textbf{Speedup}        & {\color[HTML]{F56B00} Medium}                                 & {\color[HTML]{366226} Good}                                     & {\color[HTML]{F56B00} Medium}                            & {\color[HTML]{366226} \textbf{Excellent}}                                           \\
\cmidrule(lr){2-5}
\cellcolor[HTML]{F1F1F1}\textbf{Energy reduction} & {\color[HTML]{366226} Good}                                   & {\color[HTML]{F56B00} Medium}                                   & {\color[HTML]{DF0D12} Bad}                               & {\color[HTML]{366226} \textbf{Good}}                                                \\
\cmidrule(lr){2-5}
\cellcolor[HTML]{F1F1F1}\textbf{Memory reduction}  & {\color[HTML]{366226} Good}                                   & {\color[HTML]{DF0D12} Bad}                                      & {\color[HTML]{F56B00} Medium}                            & {\color[HTML]{366226} \textbf{Good}}                                                \\
\cmidrule(lr){2-5}
\cellcolor[HTML]{F1F1F1}\textbf{Accuracy}          & {\color[HTML]{DF0D12} Bad}                                    & {\color[HTML]{F56B00} Medium}                                   & {\color[HTML]{366226} Good}                              & {\color[HTML]{366226} \textbf{Good}}                                                \\ \bottomrule
\end{tabular}
}
\end{table}

%% file: docs/3-motivation.tex
\section{Basic Implementation and Challenges}
\label{sec:challenges}

\begin{figure}
    \centering
    \includegraphics[width=1\linewidth]{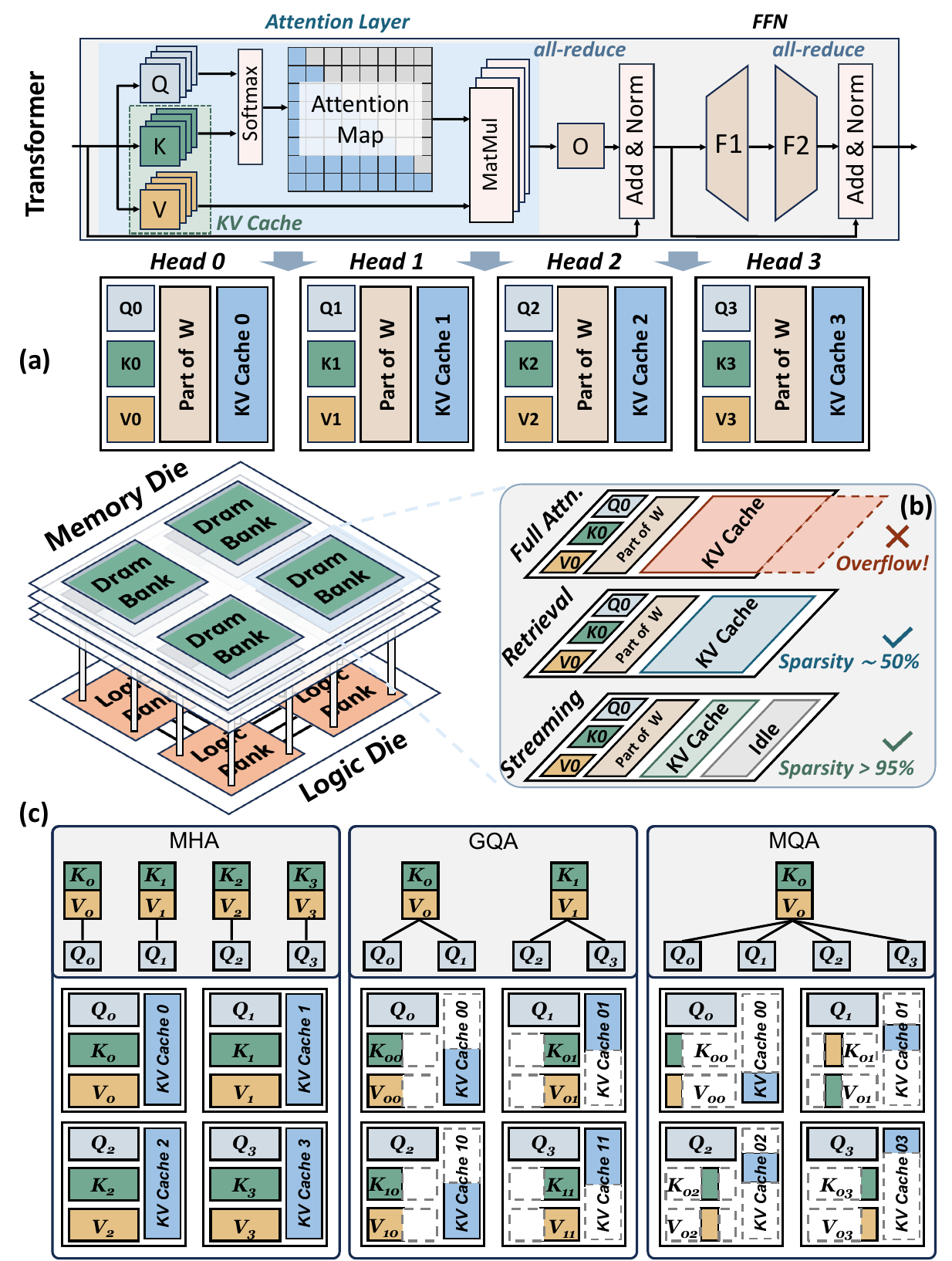}
    \caption{(a) The hybrid-bonding architecture comprises stacked memory dies on top and a logic die at the bottom, featuring distributed memory. The Basic LLM implementation on HB architectures utilizes head parallelism. (b) Full attention may lead to an overflow in the KV cache. When using sparse attention, it is observed that the sparsity varies across different heads. (c) Implementation methods of MHA, GQA, and MQA.}
    \label{fig:HBLLM}
    \vspace{-10pt}
\end{figure}


\subsection{Basic LLM Implementation on HB Architectures}
In a typical deployment of HB architectures, the distributed memory demands careful management to minimize the communication overhead.
Fortunately, in transformer-based LLM, interactions among heads are little, therefore we can just allocate different attention heads to different banks (i.e. head parallelism), as shown in Figure \ref{fig:HBLLM}(a). 
In head parallelism, only two all-reduce operations are required between banks, significantly reducing the overall communication cost.

Apart from regular multi-head attention (MHA),
group-query attention (GQA) \cite{Ainslie2023GQATG} and multi-query attention (MQA) \cite{Shazeer2019FastTD} are proposed to alleviate the memory demand of MHA
as illustrated in Figure \ref{fig:HBLLM}(c). 
Different attention mechanisms are essentially the same as we focus on the KV head.
Specifically, when centering around the KV head, once the KV cache is properly mapped and handled, the remaining weight modules can be allocated to the corresponding group of banks by tensor parallelism.
However, implementing LLMs on the HB architecture on long contexts still faces the following challenges as shown in Figure \ref{fig:overview}.





\subsection{Challenges}
\textbf{\textit{Challenge 1: Enormous KV cache overhead motivates the algorithm-hardware co-design.}}
When handling long contexts, LLMs suffer from the increasing computation and memory requirements (including both memory capacity and bandwidth) associated with the KV cache, which becomes the prominent bottleneck for long-context LLM inference and exceeds the storage capacity of HB \cite{Luohe2024KVCache} as illustrated in Figure \ref{fig:overview}.
This challenge motivates us to leverage the inherent attention sparsity to compress the KV cache size.
However, existing sparse attention methods either suffer from high accuracy degradation \cite{Xiao2023EfficientSL} or introduce extra computation costs \cite{Chen2024ArkValeEG,Tang2024QuestQS}.
Meanwhile, recent studies observe that sparsity varies across different heads \cite{Xiao2024DuoAttentionEL,Fu2024NotAH}, providing us with larger optimization space for more flexible and extreme KV cache compression.
These observations motivate us to co-design algorithm and hardware, including
1) an accurate head-wise sparse attention algorithm to achieve high sparsity without compromising accuracy; and
2) specialized hardware modules to better leverage the sparse attention with heterogeneous sparsity across heads.

\textbf{\textit{Challenge 2: Varying sparsity in different heads leads to workload imbalance between banks.}}
As described above, sparsity varies across different heads, shown in Figure \ref{fig:HBLLM}(b). 
Instead of using uniform sparsity across heads, we apply different sparse patterns for different heads. 
However, such sparsity heterogeneity among heads leads to critical synchronization bottlenecks due to data dependencies - all heads must await completion of the most computationally intensive operation before progressing to subsequent layers. 
The challenge can be more severe due to the distributed memory architecture of HB architecture \cite{Yue2024ExploitingSO,Niu2022184QPSW6L,Fujun2020ASE,Wuu20223DVT}, which fundamentally differs from conventional unified memory systems.
Hence, how to effectively balance workload distribution across memory-distributed banks for the hybrid sparsity pattern is a key question.



    

\textbf{\textit{Challenge 3: Mismatch between the heterogeneous HB architecture and attention heads.}}
One-to-one mapping between attention heads and banks
is not generally compatible with real-world diverse LLM structures since
they possess various structures.
For example, models like Vicuna-13B (40 KV heads)\cite{vicuna2023}, LLaMA-2-7B (32 KV heads)\cite{Touvron2023Llama2O}, and DeepSeek-V2-Lite (16 KV heads)\cite{deepseekv2} exhibit diverse number of heads, 
while HB architectures adopt distinct array configurations like 4$\times$4 or 4$\times$6\cite{Yue2024ExploitingSO,Niu2022184QPSW6L,Fujun2020ASE,Wuu20223DVT}. 
This necessitates algorithm-hardware mapping algorithms to adapt any number of KV heads to any HB architecture to solve the mismatch between memory banks and attention heads.
Furthermore, varying sparsity across attention heads necessitates strategic grouping before load balancing. 
Heads must be grouped and mapped to bank arrays in a way that enables computationally intensive heads to redistribute workloads to less demanding ones within the same group. Poor grouping and mapping methods can lead to significant communication overhead and waste of computing resources.


    

%% file: docs/4-method.tex
\begin{figure}
    \centering
    \includegraphics[width=1\linewidth]{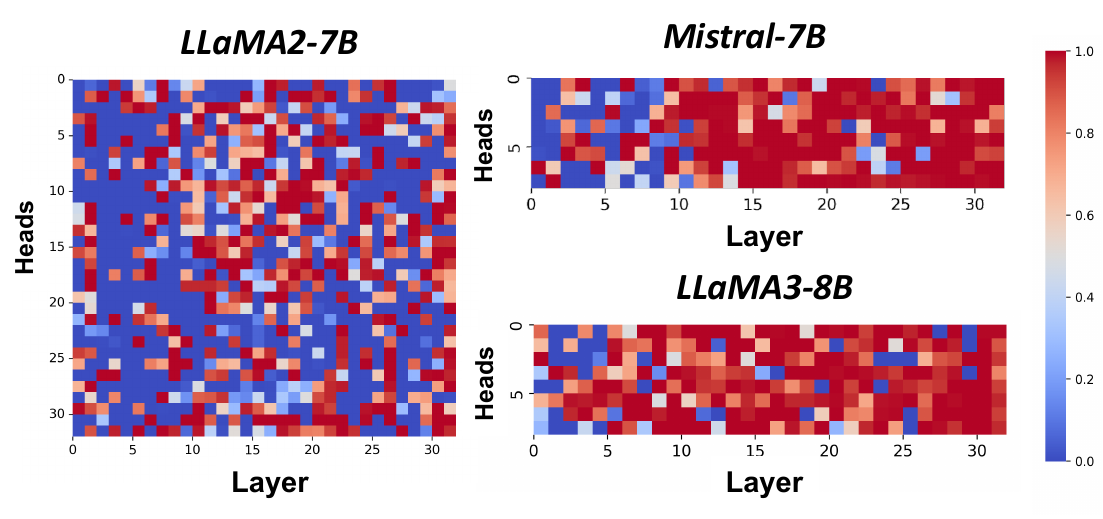}
    \caption{Attention sparsity across heads of multiple LLMs.
    Colder color indicates higher sparsity.
    }
    \label{fig:heatmap}
    \vspace{-12pt}
\end{figure}

\section{Methodology}

\subsection{Hybrid Sparse Attention with Algorithm-Hardware Co-Design}
\label{method1}

\begin{figure*}[!tb]
    \centering
    \includegraphics[width=1\linewidth]{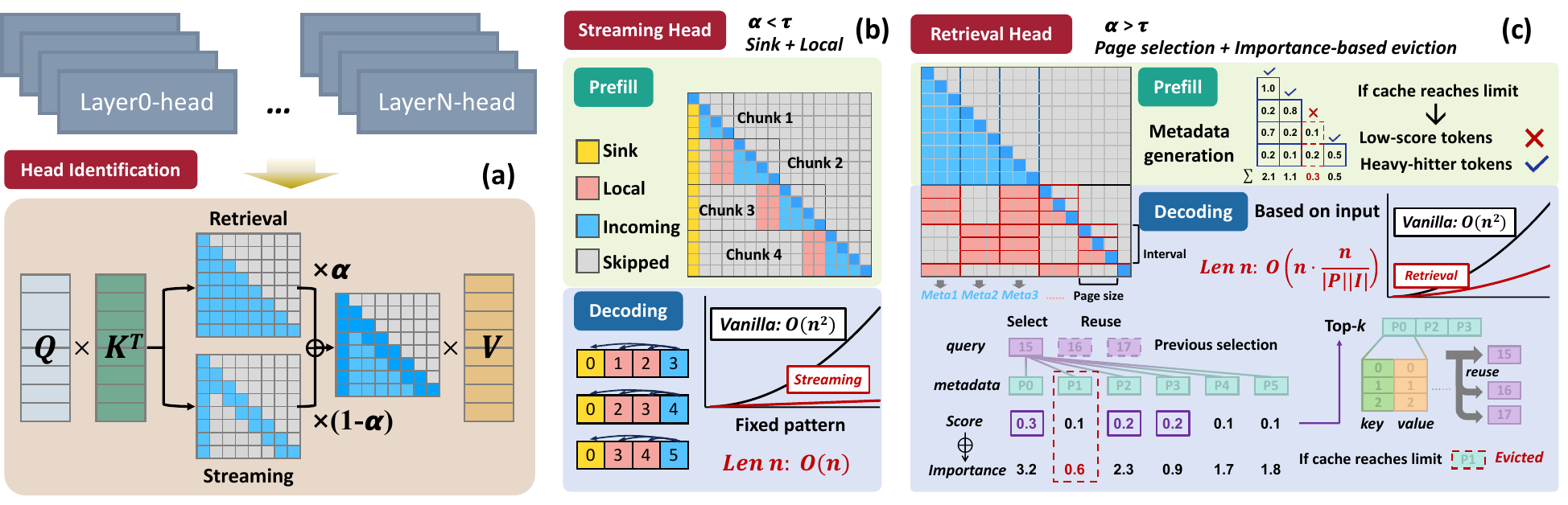}
    \caption{Hybrid static-dynamic sparse attention algorithm. \textbf{(a) Head Identification:} an optimization-based method is employed to determine the retrieval and streaming head, where the gating parameter $\alpha$ is the only trainable parameter. \textbf{(b) Streaming Head:} static sparse attention retains sink and local tokens. \textbf{(c) Retrieval head:} dynamic sparse attention utilizes page selection based on the current query.
    }
    \label{fig:pruning}
    \vspace{-10pt}
\end{figure*}


Transformer-based LLMs exhibit different sparse patterns in different heads \cite{Xiao2024DuoAttentionEL,Yang2025LServeEL,Ge2023ModelTY,Fu2024NotAH}.
Figure \ref{fig:heatmap} shows the sparse patterns varies across different heads and layers.
As shown in challenge 1 in Figure \ref{fig:overview}, some heads dynamically attend to the tokens based the the query while other heads mainly attend to the local and the first few tokens (termed as sink tokens) as observed in \cite{Xiao2024DuoAttentionEL}.
However, existing works set uniform sparse ratio for different heads, which is not sub-optimal for both efficiency and accuracy \cite{Zhang2023H2OHO,Tang2024QuestQS,Chen2024ArkValeEG}. 
Hence, to enable more flexible sparse pattern, we follow \cite{Xiao2024DuoAttentionEL,Yang2025LServeEL} and define two kinds of heads as
\begin{itemize}
    \item \textbf{Streaming head:} the attention pattern exhibits a fixed pattern where the tokens mainly attend to the local and sink tokens.
    \item \textbf{Retrieval head:} the attention pattern exhibits a dynamic pattern where the tokens attend to different tokens based on the current query.
\end{itemize}

Following \cite{Xiao2024DuoAttentionEL}, for streaming heads, we directly use the local and sink tokens for sparse attention computation.
For the retrieval heads, directly using all tokens for attention computation like \cite{Xiao2024DuoAttentionEL} is inefficient.
Therefore, we further develop a page-based dynamic selection method for retrieval heads to minimize the overhead of attention computation.

\subsubsection{Head Identification}

As shown in Figure \ref{fig:pruning}(a), we follow \cite{Xiao2024DuoAttentionEL} and adopt an optimization-based method to identify two kinds of heads using the gating parameter $\alpha$.
During training, we regard the retrieval head as full attention. 
The optimization can be formulated as
\begin{equation*}
    \mathrm{Attn}_{i,j} = \alpha_{i,j}\cdot \mathrm{Full\_Attn} + (1-\alpha_{i,j})\cdot \mathrm{Streaming\_Attn},
\end{equation*}
where $\alpha_{i,j}\in [0,1]$ denotes the gating parameter for the $i$-th layer and $j$-th head. At beginning, $\alpha$'s are initialized to 1.

\subsubsection{Static Sparsity of Streaming Heads}


As illustrated in Figure \ref{fig:pruning}(b), streaming heads focus on the local and sink tokens. Since the sparse pattern is pre-determined and fixed during the whole inference process, streaming heads dramatically reduce the memory usage and memory bandwidth demand.

\subsubsection{Dynamic Sparsity of Retrieval Heads}



retrieval heads dynamically select relevant tokens based on the current query.
Inspired by \cite{Tang2024QuestQS,Chen2024ArkValeEG,zeng2025mpcache},
we group contiguous KV caches into pages for higher selection efficiency.
We use the minimum and maximum boundary of each page to approximate the relevance score \cite{Tang2024QuestQS,Chen2024ArkValeEG}.
Each page can be represented using metadata $\tau$ where
\begin{equation*}
    \tau_{\max} = \max_{k\in K_p} k,~~\tau_{\min} = \min_{k\in K_p} k,
\end{equation*}
where \(K_p\) denotes all the keys within a page, and $\min, \max$ denote element-wise operations.

Dynamic selection can be depicted as a two-step pipeline:
\begin{itemize}
    \item \textbf{Relevance score computation:} first compute the relevance score between the current query and each page's metadata using $\max(q\cdot \tau_{\min}, q\cdot \tau_{\max})$, where $q$ denotes the query.
    \item \textbf{Top-$k$ selection:} then select the top-$k$ most relevant pages to compute the following sparse attention.
\end{itemize}
To further minimize the overhead of dynamic selection, we share the page selection for near queries \cite{Yang2025LServeEL}.

\textbf{Memory consideration.} Considering the memory limitation and hardware constraint, the increasing KV cache size during the decoding process is not friendly. 
Hence, we keep and update the accumulated attention scores for pages based on the computed relevance score \cite{Zhang2023H2OHO} at each step, and then,
evict one page with the lowest accumulated attention score when the KV cache size reaches the memory budget.

\subsubsection{Hardware Co-Design for Hybrid Sparse Attention}

Streaming heads only attend to the sink and local tokens while retrieval heads are more complex:
metadata needs first to be loaded into the logic die for $\min/\max$ computation, followed by the top-$k$ selection from the KV cache.
Then, the selected KV cache is loaded into the logic die for attention computation.
Following \cite{Xiao2023EfficientSL}, retrieval heads also attend to sink and local tokens.
As shown in Figure \ref{fig:codesign}, we implement the following designs on hardware:

\begin{figure}
    \centering
    \includegraphics[width=1\linewidth]{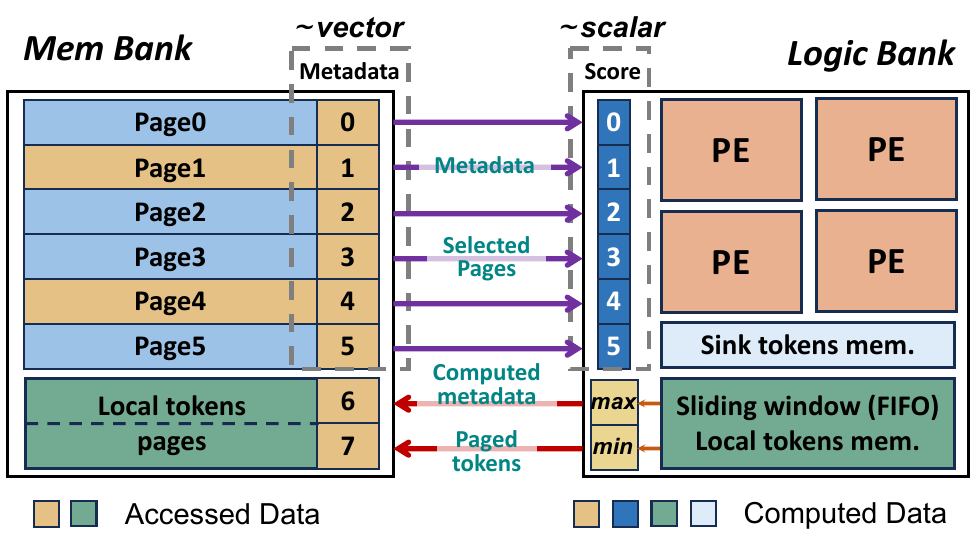}
    \caption{Hardware co-design for sparse attention.}
    \label{fig:codesign}
    \vspace{-12pt}
\end{figure}

\begin{itemize}
    \item \textbf{Sink and local tokens in the logic bank.} All sink tokens and a subset of local tokens are stored on the logic die, as they are accessed by all attention heads and require relatively smaller space. Given that the number of local tokens typically ranges from 128 to 512, the majority are still stored on the memory die. The importance score is also stored in the logic bank. After each attention operation, the attention score of the corresponding page is accumulated into the importance score.
    \item \textbf{Paged tokens and metadata in the memory bank.} We implement a sliding window for local tokens using a FIFO. For streaming head, popped tokens are sent to memory die and the least recent tokens are discarded. For retrieval head, popped paged tokens pass through specialized min and max units to generate metadata, which is then sent to the memory die. If the memory die is full, the least important page is discarded based on its importance score. We store the metadata corresponding to each page. Although the metadata is relatively small, it grows with the sequence length increasing, making it unsuitable for placement on logic die. Only the local tokens, metadata, and the top-$k$ relevant pages need to be accessed.
\end{itemize}

\subsection{Cross-Head Load Balancing with Interleaved Memory-Compute Co-Placement}
\label{method2}


\subsubsection{Load Balance between Different Types of Heads}
As discussed in Section \ref{method1}, the hybrid sparse pattern across different heads leads to load imbalance. 
The naive one-to-one uniform mapping of all heads leads to sub-optimal hardware utilization as shown in challenge 2 in Figure \ref{fig:overview}. 
To address this problem, after co-locating retrieval heads with neighboring
streaming heads as introduced in Section \ref{method3}, we evenly distribute KV cache operations across associated memory banks within each tile, effectively balancing the heavier computational load from retrieval heads. 
This design ensures efficient hardware resource utilization by leveraging underutilized capacity in adjacent banks through workload partitioning.

\subsubsection{Memory-Compute Co-Placement for Multi-Bank KV Cache}

Load balancing introduces a new challenge, i.e., \textit{how to effectively distribute KV cache operations across multiple memory banks?} 
Unlike the unified memory architecture of GPU, HB technology employs physically distributed memory where each bank maintains an independent storage resource. 
These challenges motivate us to simultaneously optimize both computation and memory allocation.
Therefore, we propose the memory-compute co-placement that maps each token's KV cache computation and their associated storage to the same bank.
This architectural principle ensures localized processing of all KV cache operations for assigned tokens. 
When computing Softmax, we efficiently handle cross-bank communication following FlashAttention \cite{Dao2022FlashAttentionFA}.
As long as the KV cache needs to be distributed across banks, we apply the memory-compute co-placement strategy.

\subsubsection{Interleaved KV Cache Storage between Banks}

\begin{figure}
    \centering
    \includegraphics[width=1\linewidth]{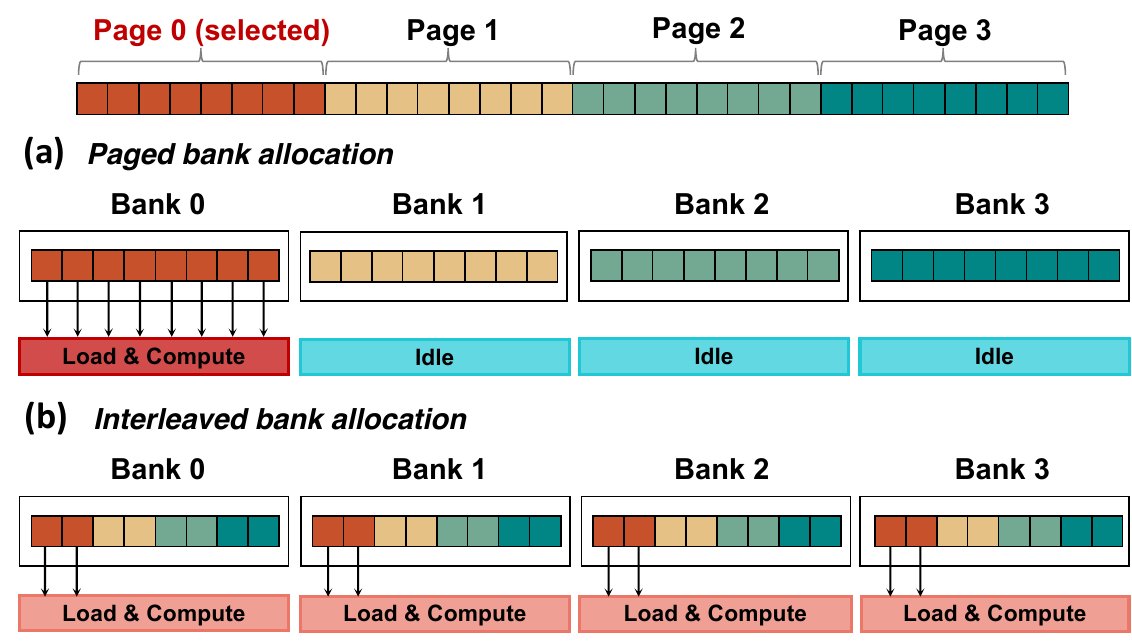}
    \caption{Comparison of workload across banks between paged bank allocation and interleaved bank allocation.}
    \label{fig:interleave}
    \vspace{-10pt}
\end{figure}
  
Dynamic selection introduces a fundamental challenge, i.e., \textit{the unpredictable distribution of top-$k$ KV cache pages across different inputs creates a potential workload imbalance.} 
As shown in Figure \ref{fig:interleave}(a), naively allocating one page in one bank can face computational imbalance when selected pages concentrate in specific banks.
To eliminate this imbalance induced by dynamic selection, we proposed an interleaved solution.
Specifically, the tokens within one page are evenly distributed across multiple banks as illustrated in Figure \ref{fig:interleave}(b),
ensuring balanced memory access and computational workloads across all banks, regardless of input-specific paged selection. 
By ensuring proportional workload distribution for arbitrary selection, 
our interleaving method addresses the inherent unpredictability of dynamic sparse attention.



\subsection{Adaptive Heterogeneous Mapping with Communication-Minimal Tiling}
\label{method3}


As mentioned in Section \ref{sec:challenges}, the mismatch between HB architecture and attention heads, as well as hybrid sparse patterns, motivates us to propose the adaptive mapping strategy.

\subsubsection{Mapping Heads of Different Amount}
We will proceed with different cases of different size relationship between $n_h$ and $n_b$:

\textbf{(a) $\boldsymbol{n_h = n_b}$ or $\boldsymbol{n_b}$ is divisible by $\boldsymbol{n_h}$}: This is the most basic case where each head can be directly assigned $n_b/n_h$ banks as a group and tensor parallelism will be implemented within the group. And our target is to transform any other case into this basic one.

\textbf{(b) $\boldsymbol{n_h > n_b}$}: the heads can be decomposed into $\lceil n_h / n_b \rceil$ disjoint subsets $\{n_{h1}, n_{h2}, \ldots, n_{hk}\}$, where each subset satisfies $|n_{hi}| \leq n_b$. These subsets are then executed sequentially through a pipeline:
$
n_{h1} \rightarrow n_{h2} \rightarrow \cdots \rightarrow n_{hk}.
$
Hence, this case can be reduced to $n_h \leq n_b$

\textbf{(c) $\boldsymbol{n_h < n_b}$ and $\boldsymbol{n_b}$ is not divisible by $\boldsymbol{n_h}$}:  A greedy decomposition strategy is proposed to partition \(n_h\) into a sum of \(l\) distinct divisors of \(n_b\) (denoted as \(\{n_{h1}', n_{h2}', \ldots, n_{hl}'\}\)) such that
\[
n_h=\sum_{i=1}^l n_{hi}'\quad\mathrm{with}~ n_b~\mathrm{mod}~ n_{hi}' = 0, \forall i \in \{1, 2, \ldots, l\}.
\]

The subsets are also executed sequentially. To minimize the number of pipeline stages, the decomposition prioritizes larger divisors first every time (greedy strategy). For each subset \(n_{hi}'\), just like case(a), the banks are partitioned into \(n_{hi}'\) groups, with each group containing $n_b/n_{hi}'$ banks.

\begin{figure}
    \centering
    \includegraphics[width=1\linewidth]{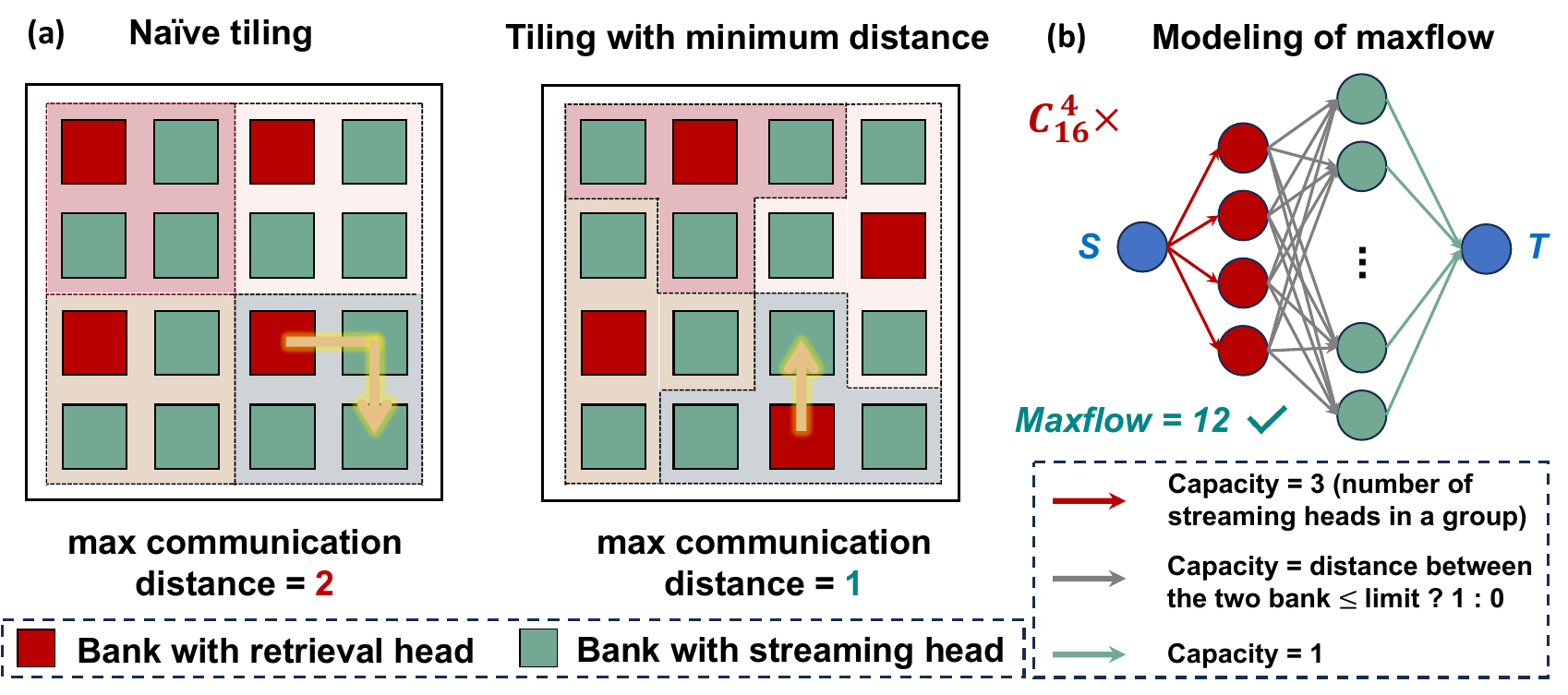}
    \caption{Different tiling strategy on a $4\times4$ mesh-like NoC including 4 banks with retrieval head and 12 banks with streaming head.}
    \label{fig:tiling}
    \vspace{-10pt}
\end{figure}

\subsubsection{Mapping Heads of Different Sparsity}
As mentioned in Section \ref{method2}, we are supposed to balance the workload of different kinds of heads. To map heads of different sparsity, we introduce \textit{tiling} - heads are partitioned into tiles where each tile contains a combination of retrieval and streaming heads. 
Workload will be shared within the tile by memory-compute co-placement illustrated in Section \ref{method2}. 

Denote the number of retrieval heads as $n_r$, streaming heads as $n_s$, the number of tiles as $t$, and the set of tiles $T$ as $\{T_0, T_1, \dots, T_{t-1}\}$,
the tiling problem can be formulated as
\begin{align*}
    &\min~\mathrm{MaxDist}(\mathrm{Loc}_r, \mathrm{Loc}_s) \\
    &\mathrm{where}~\mathrm{Loc}_r, \mathrm{Loc}_s \in T_i~(i=0, \dots, t-1), \\
    &\mathrm{s.t.}~t = \min(n_r, n_s), |T_i| \leq \lceil (n_s+n_r)/t \rceil,
\end{align*}
where $\mathrm{Loc}_r, \mathrm{Loc}_s$ denote the physical location of retrieval head and streaming head, $\mathrm{MaxDist}$ denotes the maximum distance between the retrieval head and streaming head within each tile, and $|T_i|$ denotes the tile size of the $i$-th tile.

In the formulation, we balance the number of retrieval heads and streaming heads
through the following constraints:
\ding{182} setting number of tiles $t$ to the smaller value between $n_r$ and $n_s$, $t = \min(n_r, n_s)$; 
\ding{183} making the size of each tile not exceed $\lceil (n_s+n_r)/t \rceil$.
Therefore, we ensure the retrieval heads and streaming heads are equally arranged in each tile.

Our objective is to minimize the maximum distance between two kinds of heads since the workload share brings communication overhead.
Figure \ref{fig:tiling}(a) shows the difference between naive tiling without considering the communication distance and tiling with communication distance minimized.
We model and solve this problem as a network maxflow problem \cite{Ford1956MaximalFT} as illustrated in Figure \ref{fig:tiling}(b).

%% file: docs/5-experiments.tex
\section{Experiments}

\subsection{Experimental Setups}

\subsubsection{Hardware Setups}

We develop a cycle-level simulation framework to model the HB-based accelerator, integrating logic and memory dies vertically. 
The memory sub-system employs a multi-stack architecture consisting of 4 layers, with a macro bandwidth of 256 bits \cite{Yue2024ExploitingSO} and access energy 0.88pJ/bit \cite{Niu2022184QPSW6L}. 
Following \cite{Tsmc2021cim}, we build and model the General Matrix Multiplication (GEMM) accelerator consisting of 16 DCIM macros, each with 900 GOPS throughput and 24 TOPS/W energy efficiency. 
Detailed specifications are presented in Table \ref{tab:specification}.

\begin{table}
    \centering
    \caption{Hardware specification of \method~\cite{Yue2024ExploitingSO,Niu2022184QPSW6L,Tsmc2021cim}.}
    \label{tab:specification}
    \begin{tabular}{m{0.25\linewidth}>{\raggedright\arraybackslash}m{0.6\linewidth}}
        \Xhline{1pt}
         \makecell[l]{\textbf{Logic Die}} & \makecell[l]{\textbf{Specification}} \\
        \Xhline{1pt}
        Technology & 22nm \\
        Area & $25mm\times28mm$ \\
        Throughput& 900GOPS$\times$16/bank@int8 \\
        Energy Efficiency& 24 TOPS/W@int8 \\
        SRAM Cache&8$\times$128KB/bank \\
        NoC & $4\times4$ 2-d Mesh, 256 bits bandwidth\\
        Voltage & VDD=0.7V\\
        \Xhline{1pt}
         \makecell[l]{\textbf{Memory Die}} &  \makecell[l]{\textbf{Specification}}\\
        \Xhline{1pt}
        Area & $25mm\times28mm$\\
        Capacity & 32GB: 32MB/macro, 16macros/bank, 16banks/layer, 4 layers stacked\\
        Bandwidth & 256bits/4Macros$\cdot$cycle \\
        Access Energy & 0.88pJ/bit\\
        Refresh Period& $<$85C:32ms; $<$95C:16ms;     $<$105C:8ms; $<$115C:4ms\\
        Frequency & 400MHz \\
        Voltage & VDD=1.2V\\ 
        \Xhline{1pt}
         
    \end{tabular}
\end{table}

\subsubsection{Software Setups}

For LLM architectures, We evaluated the end-to-end performance on Mistral-7B, LLaMA2-7B, and LLaMA3-8B (quantized to 8-bit weights, 8-bit activations, and 8-bit KV)\cite{Lin2024QServeWQ}, which are popular models in edge scenarios. 
For evaluation datasets, the accuracy evaluation is conducted on LongBench \cite{longbench} (up to 22k words) and Needle-in-a-Haystack (NIAH) \cite{kamradt2024llmtest} benchmarks.

\begin{figure}
    \centering
    \includegraphics[width=1\linewidth]{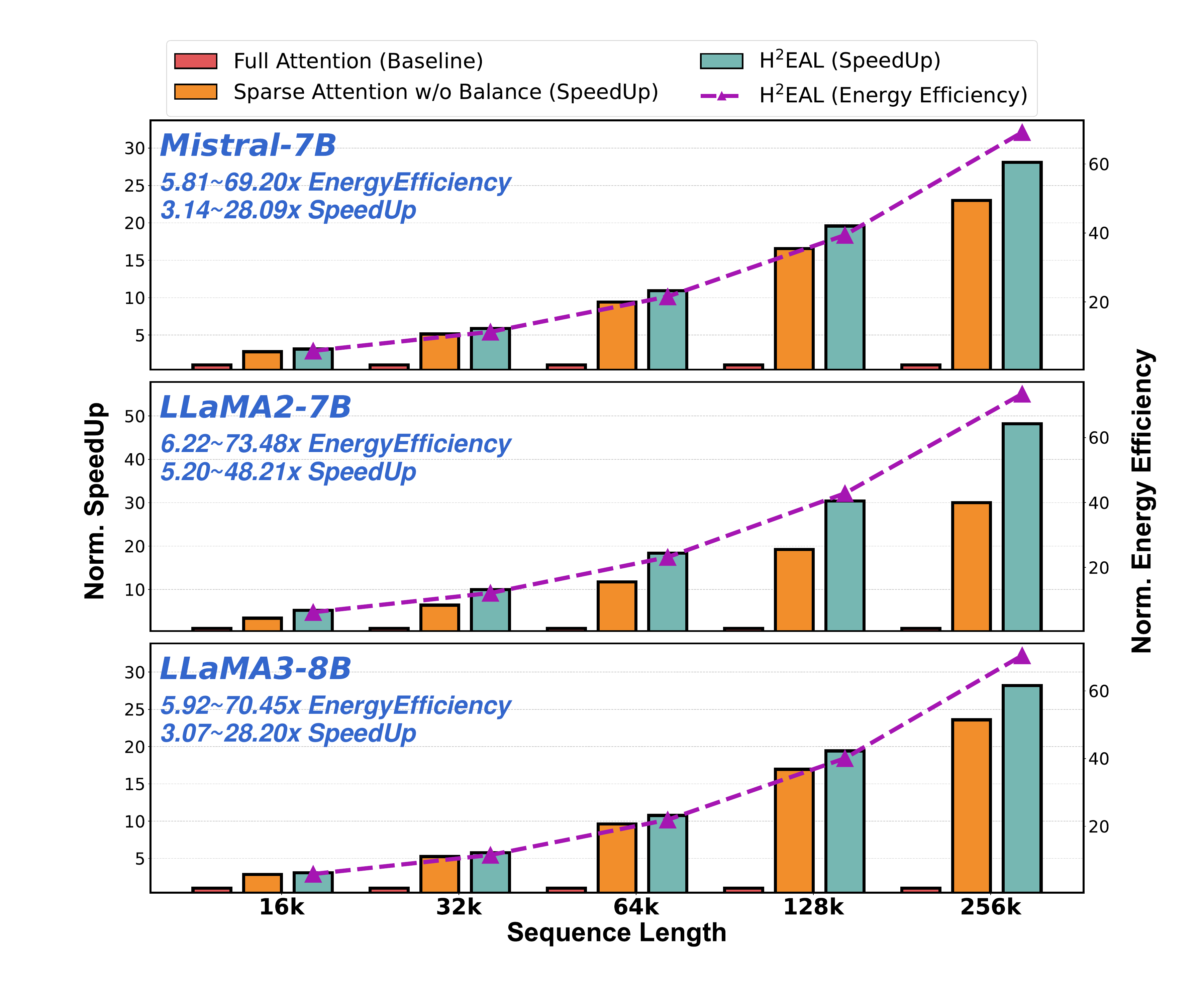}
    \caption{Normalized attention speedup and energy efficiency of full attention, sparse attention w/o balance and \method, on Mistral-7B, LLaMA2-7B, and LLaMA3-8B.}
    \label{fig:performance2}
\end{figure}

\begin{figure}
    \centering
    \includegraphics[width=1\linewidth]{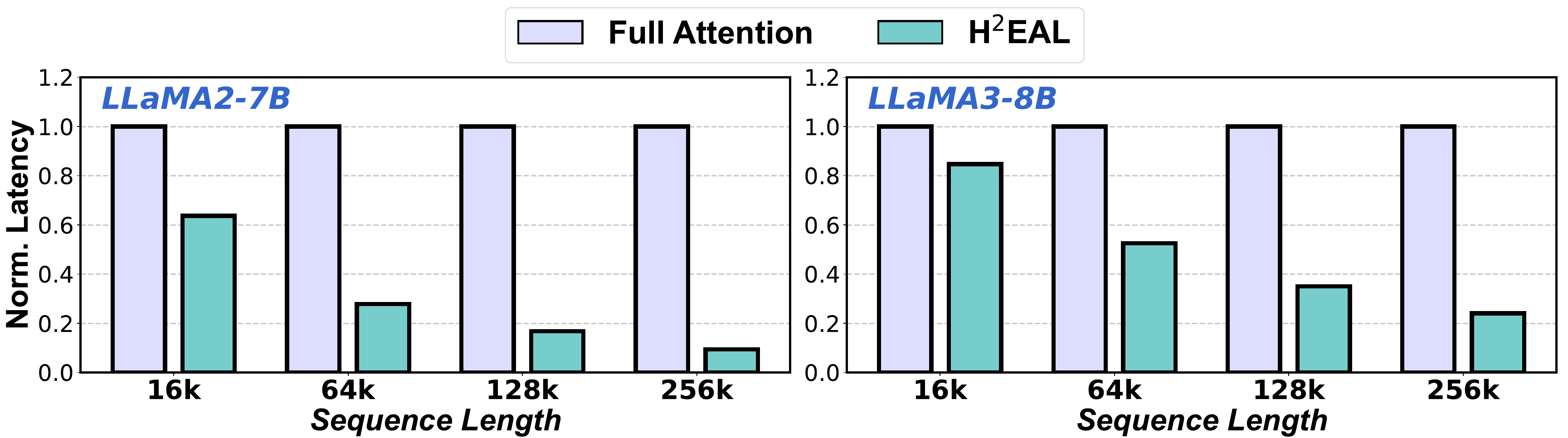}
    \caption{Normalized end-to-end latency comparison.}
    \label{fig:end2endnorm}
    \vspace{-10pt}
\end{figure}

\subsection{Performance Evaluation}

We implement the baselines 1) full attention; 2) sparse attention without tile balancing, and compare these methods with our \method.
For \method, we set the static sparsity (i.e. the proportion of streaming heads) to 0.5, the total selection length for the retrieval head to 4k, and the page size to 32.

\subsubsection{Micro-benchmark for Attention Evaluation}

Figure \ref{fig:performance2} illustrates the normalized attention speedup and energy efficiency for different decoding sequence lengths. 
As observed, we can make the following conclusions with the sequence length of 256k:
\begin{itemize}
    \item Compared to full attention, \method~achieves 28.09$\times$, 48.21$\times$, and 28.20$\times$ speedup on Mistral-7B, LLaMA2-7B and LLaMA3-8B, respectively;
    \item Compared to sparse attention without balance, \method~achieves 1.221$\times$, 1.605$\times$, and 1.195$\times$ speedup, on Mistral-7B, LLaMA2-7B and LLaMA3-8B, respectively;
    \item For energy efficiency, compared to full attention, \method~achieves 69.20$\times$, 73.48$\times$, and 70.45$\times$ improvement on Mistral-7B, LLaMA2-7B and LLaMA3-8B, respectively.
\end{itemize}

\subsubsection{End-to-end Evaluation}

Table \ref{tab:end2endperformance} presents the end-to-end throughput and energy efficiency of full attention and our \method~on LLaMA2-7B and LLaMA3-8B. 
Results show that with the sequence length of 256k, \method~achieves 430.8 and 469.7 tokens/s end-to-end throughput for LLaMA2-7B and LLaMA3-8B, respectively. 
Energy efficiency of \method~at 256k is 23.20 and 25.83 tokens/J, respectively, which corresponds to 12.21$\times$ and 4.27$\times$ improvement to full attention.
Figure \ref{fig:end2endnorm} shows the normalized latency with different sequence lengths. 
As observed, \method~achieves 10.56$\times$ and 4.15$\times$ speedup compared to full attention with the sequence length of 256k.

\subsubsection{Ablation Study of Balancing Optimization}

\begin{figure}
    \centering
    \includegraphics[width=1\linewidth]{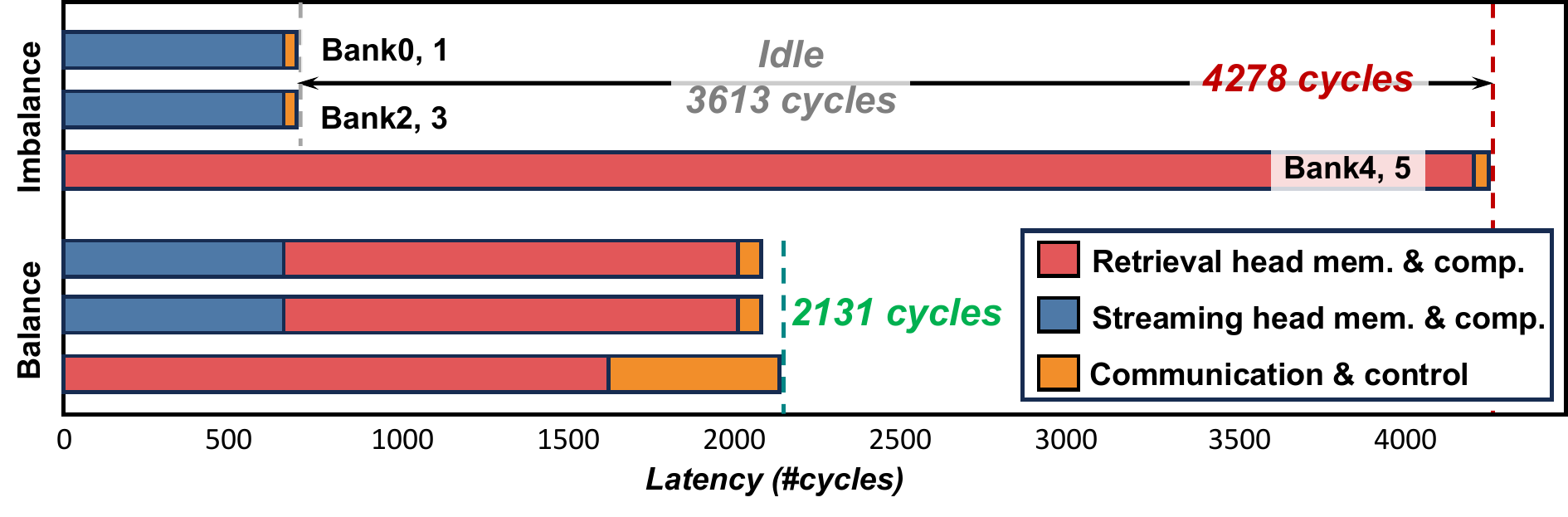}
    \caption{Latency breakdown before and after balancing. The example uses three heads within a tile of one attention layer on LLaMA3-8B and a sequence length of 12k.}
    \label{fig:balance}
\end{figure}

To dive into the effectiveness of balancing optimization, Figure \ref{fig:balance} depicts the latency breakdown before and after balancing.
The unbalanced situation brings 3613 cycles of idle on banks 0-4 while the balanced one effectively eliminates it.
In this example, balancing optimization offers a speedup of 2.01$\times$.



\begin{table}
    \centering
    \caption{End-to-End throughput and energy efficiency of LLaMA2-7B and LLaMA3-8B for long-context inference.}
    \label{tab:end2endperformance}
    \begin{tabular}{cccccc}
        \Xhline{1pt}
        \multicolumn{2}{c}{\textbf{Model}} & \multicolumn{2}{c}{LLaMA2-7B} & \multicolumn{2}{c}{LLaMA3-8B} \\
        \multicolumn{2}{c}{\textbf{Sequence Length}} &  64k & 256k & 64k & 256k \\
        \Xhline{0.5pt}
        Throughput & Full Attention & 127.9 & 40.8 & 253.4 & 113.1\\
        (tokens/s) & \method & 459.5 & 430.8 & 482.1 & 469.7\\
        \Xhline{0.5pt}
        Energy Efficiency & Full Attention& 6.32 & 1.90 & 14.69 & 6.05\\
        (tokens/J) & \method & 24.00 & 23.20 & 26.10 & 25.83\\
        \Xhline{1pt}
    \end{tabular}
\end{table}

\begin{table}[!tb]
\caption{Accuracy evaluation of \method~on LongBench 21 datasets with LLaMA3-8B model.}
\label{tab:longbench}

\resizebox{\columnwidth}{!}{

\begin{tabular}{@{}ccc|ccc@{}}
\toprule
\textbf{Dataset}       & \textbf{Full}  & \textbf{\method}  & \textbf{Dataset}           & \textbf{Full}           & \textbf{\method}           \\ \midrule
2WikiMQA        & 29.19 & 31.13 & PassageCount        & 2.00           & 1.09           \\
DuReader        & 30.36 & 30.33 & PR-en & 77.00          & 77.05          \\
GovReport       & 34.52 & 32.35 & PR-zh & 63.01          & 60.59          \\
HotpotQA        & 42.25 & 41.21 & Qasper              & 28.59          & 27.86          \\
LCC             & 34.81 & 37.89 & QMSum               & 24.15          & 24.74          \\
LSHT            & 38.00 & 31.50 & RepoBench-P         & 37.26          & 37.11          \\
MultiNews       & 27.74 & 27.53 & SAMSum              & 41.80          & 40.26          \\
MultiFieldQA-en & 50.99 & 50.15 & TREC                & 71.50          & 73.00          \\
MultiFieldQA-zh & 48.91 & 47.99 & TriviaQA            & 87.59          & 84.88          \\
MuSiQue         & 22.12 & 23.50 & VCSUM               & 11.04          & 10.19          \\
NarrativeQA     & 26.30 & 26.04 & \textbf{Average}    & \textbf{39.48} & \textbf{38.88} \\ \bottomrule
\end{tabular}
}
\end{table}

\subsection{Accuracy Evaluation}
\label{experiment-2}




We evaluate the accuracy of our proposed \method~on the LLaMA3-8B \cite{Lin2024QServeWQ} across the 21 datasets from LongBench \cite{longbench} and the Needle-in-a-Haystack (NIAH) benchmark \cite{kamradt2024llmtest}. 

\begin{figure}[!tb]
    \centering
    \includegraphics[width=1\linewidth]{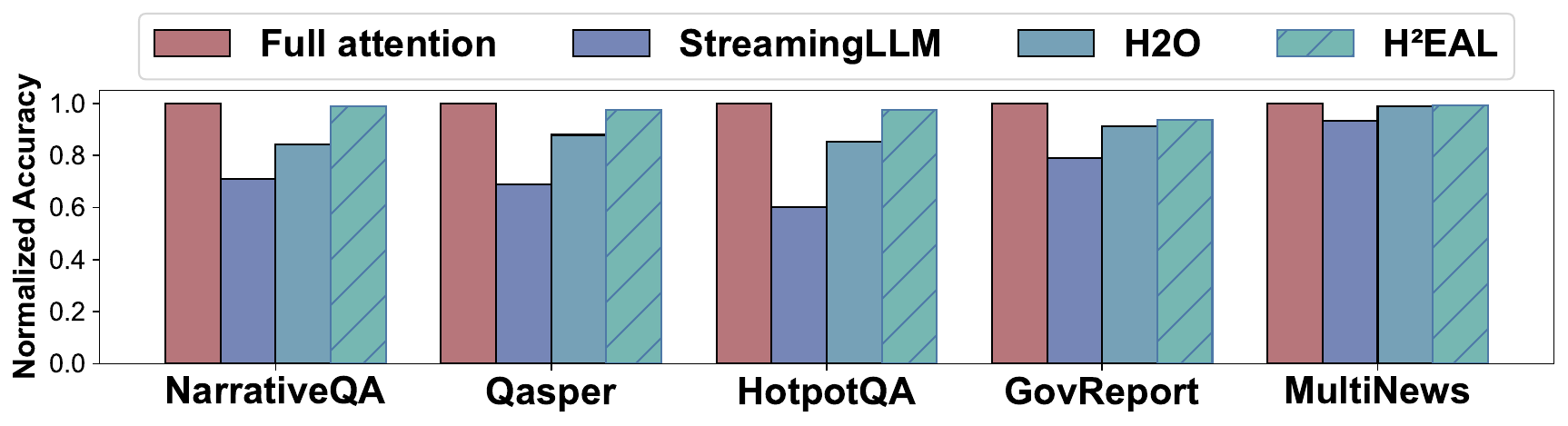}
    \caption{The accuracy comparison of \method~with full attention, StreamingLLM \cite{Xiao2023EfficientSL} and H2O \cite{Zhang2023H2OHO} across multiple datasets in LongBench \cite{longbench}.}
    \label{fig:acc_compare}
    \vspace{-5pt}
\end{figure}

\begin{figure}[!tb]
    \centering
    \includegraphics[width=1\linewidth]{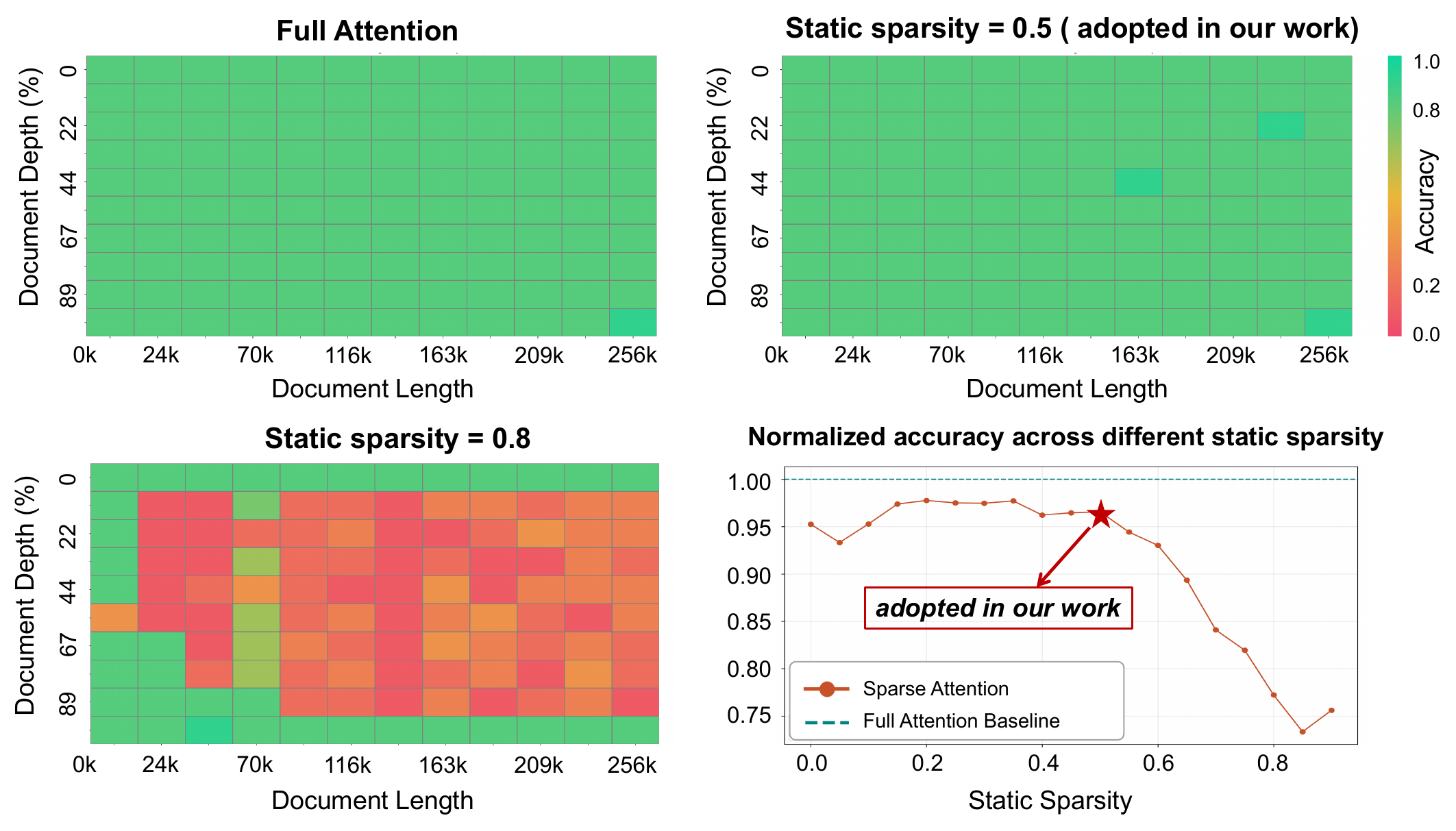}
    \caption{Ablation study on under different static sparsity. 
    The first three subfigures are Needle-In-A-Haystack.
    The x-axis denotes the length of the context (``haystack'') and the y-axis indicates the position where the ``needle'' (a short prompt) is inserted within the context.
    The last subfigure is the average accuracy on LongBench.
    }
    \label{fig:niah}
    \vspace{-10pt}
\end{figure}

\subsubsection{LongBench Result}
Table \ref{tab:longbench} demonstrates that \method~achieves comparable or even improved performance compared to full attention across the majority of datasets.
The reason why sparse attention outperforms full attention is likely due to the ability to mitigate attention noise for long contexts.
Figure \ref{fig:acc_compare} demonstrates the comparison between \method~and recent works \cite{Xiao2023EfficientSL, Zhang2023H2OHO},
and \method~consistently achieves higher accuracy than StreamingLLM \cite{Xiao2023EfficientSL} and H2O \cite{Zhang2023H2OHO} across different datasets.

\subsubsection{Sparsity Discussion}
As illustrated in Figure \ref{fig:niah}, \method~consistently achieves strong performance under varying static sparsity on NIAH.
In particular, the static sparsity of 0.5 provides a favorable trade-off, preserving accuracy while yielding significant efficiency gains. 
The curve on LongBench also
demonstrate the trade-off and robustness of \method~and highlight its effectiveness of \method~in retaining model accuracy under aggressive KV cache compression.

%% file: docs/6-conclusion.tex
\section{Conclusion}
In this paper, we proposed \method, an HB-based accelerator for efficient edge LLM inference in long-context scenarios.
\method~combines sparse attention and hardware with careful algorithm-hardware co-design.
At the algorithm level, \method~proposes an accurate and hybrid sparse attention scheme with static and dynamic sparsity.
At the hardware level, to manage the KV cache management and address the workload imbalance, \method~proposes memory-compute co-placement and adaptive heterogeneous mapping.
Extensive results demonstrate that our proposed \method~achieves $\text{5.20}\sim\text{48.21}\times$ speedup and $\text{6.22}\sim\text{73.48}\times$ energy efficiency improvement with negligible accuracy degradation on different benchmarks.

%% file: docs/7-acknowledgment.tex
\section*{ACKNOWLEDGMENTS}
This work was supported in part by NSFC under Grant 62495102, Grant 92464104, Grant 62125401, and Grant U24A20287, in part by Beijing Outstanding Young Scientist Program under Grant JWZQ20240101004, in part by Beijing Municipal Science and Technology Program under Grant Z241100004224015, and in part by 111 Project under Grant B18001.
We thank the authors of  DuoAttention\cite{Xiao2024DuoAttentionEL}, Lserve\cite{Yang2025LServeEL} for releasing their code and for helpful discussions, which facilitated our implementation.